# Time-dependent diffusion in undulating structures: Impact on axon diameter estimation

Jan Brabec[1], Samo Lasič[2], Markus Nilsson[3]

1. Lund University, Department of Clinical Sciences Lund, Medical Radiation Physics, Lund, Sweden
2. Random Walk Imaging AB, Lund, Sweden
3. Lund University, Department of Clinical Sciences Lund, Diagnostic Radiology, Lund, Sweden

**Word count:**

6387

**Sponsors/Grant numbers:**

This research was supported by the Swedish Research Council (grant no. 2016-03443), the Swedish Foundation for Strategic Research (grant no. AM13-0090), Crafoord Foundation (grant no. 20170825), and Random Walk Imaging AB (grant no. MN15).

**Up to 8 keywords:**

diffusion MRI, time-dependent, diffusion spectrum, axonal trajectories, axon diameter, restricted diffusion, undulation

## Abbreviations:

AR    - Autoregressive process

MR    - Magnetic resonance

MRI    - Magnetic resonance imaging

dMRI  - Diffusion magnetic resonance imaging

CNS   - Central nervous system

SNR   - Signal-to-noise ratio

## Abstract

Diffusion MRI may enable non-invasive mapping of axonal microstructure. Most approaches infer axon diameters from effects of time-dependent diffusion on the diffusion-weighted MR signal by modelling axons as straight cylinders. Axons do not, however, run in straight trajectories and so far, the impact of the axonal trajectory on diameter estimation has not been systematically investigated. Here, we employ a toy-model of axons, which we refer to as undulating thin-fiber model, to analyze the impact of undulating trajectories on the diffusion-time dependence represented by the diffusion spectrum. We analyze the spectrum by its height (diffusivity at high frequencies), width (half width at half maximum), and low-frequency behavior (power law exponent). Results show that microscopic orientation dispersion of the thin-fibers is the main parameter that determines the characteristics of the diffusion spectra. Straight cylinders and undulating thin-fibers have virtually identical spectra at lower frequencies. If the straight-cylinder assumption is used to interpret data from undulating thin axons, the diameter is overestimated by an amount proportional to the undulation amplitude and the microscopic orientation dispersion. At high frequencies (short diffusion times), spectra from cylinders and undulating thin-fibers differ. The spectra from the undulating thin-fibers can also differ from that of cylinders by exhibiting power law behaviors with exponents below two. In conclusion, we argue that the non-straight nature of axonal trajectories should not be ignored when analyzing dMRI data and that careful experiments may enable separation of diffusion within straight cylinders and diffusion in undulating thin-fibers.

## Graphical Abstract

We quantified characteristics of the time-dependent diffusion in a toy-model of undulating axons by analysis of the diffusion spectra. Microscopic orientation dispersion of the undulating thin-fibers was the main factor determining the spectrum. At lower frequencies, the spectra were virtually identical to those from straight cylinders.

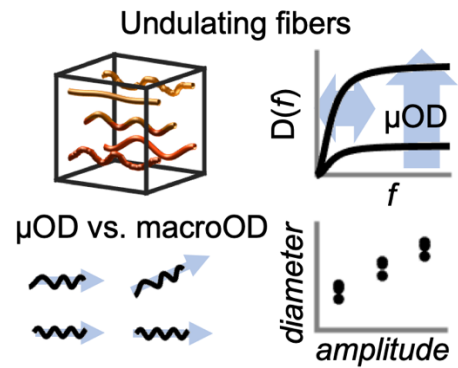

# Introduction

Tissue microstructure emerges from a finer biochemical scale and is tightly coupled with its biological function. For example, the conduction velocity through the axon is determined by its diameter, which vary in size in central nervous system between approximately 0.1–15 micrometers [1 2]. Non-invasive quantification of microstructural properties is thus appealing, and microstructural imaging with diffusion MRI has emerged as a promising technology for this purpose [3].

Quantification of the axon diameter has been a focus of many dMRI studies. Stanisz et al. analyzed dMRI data from the optic nerve with a compartment model and found the estimated parameters to reflect histological counterparts [4]. Assaf et al. used q-space imaging to show that mean square displacements of water molecules in the rat spinal cord are on the same order of magnitude as the axon diameter [5]. Later work found model-based estimates of the axon diameter distribution in the excised porcine and sciatic nerves to be in good agreement with values obtained from histology [6]. Barazany et al. reported axon diameter distributions of the rat corpus callosum estimated in-vivo [7]. Alexander et al. found an axon diameter index in the human corpus callosum in-vivo to be substantially higher than expected, but aligned with known trends [8].

In all of these studies, the axon diameter mapping was based on modeling the time-dependence of the diffusion within the axon. Analytical models for time-dependent diffusion exist for simple geometries such as cylinders, spheres, and ellipsoids [9 10]. The early model by Stanisz et al. represented axons as prolate ellipsoids, glial cells as spheres with partially permeable membranes and the extra-axonal space between as having anisotropic but time-independent (Gaussian) diffusion [4]. Later models represented axons by straight impermeable cylinders, neglected glial cells, and retained the assumption of Gaussian diffusion in the extra-axonal space [6 8 11]. Axon diameters have been represented by a single value (CHARMED [11] and ActiveAx [8]) or by a gamma distribution of diameters (AxCaliber [6]). Apart from the axon diameter, the models explain the signal also by other features of the tissue such as the density of axons [12-14], their orientation [15 16], and, in some cases, the geometry of the extra-axonal space [17 18].

The assumption that axons can be modelled as straight impermeable cylinders is today widely used. However, the validity of this assumption can be questioned [18 19]. For example, the diameter of an axon can vary along its length [20-22] and axons may feature fine morphological details such as spines, leaflets or beads [23]. Perhaps most important is that axons do not run straight [24]. Some axons exhibit sinusoidal trajectories with undulation amplitudes an order of magnitude higher than the diameter. Such undulating axons are present extra-cranially in, for example, the phrenic nerve [25] and in the cranial nerves, such as the root of the trigeminal nerve [26]. Undulations also appear to be present in at least in parts of the central nervous system, e.g. in the corona radiata, the optical nerve radiations, and in the corpus callosum [24 27]. This is important because it could lead to overestimated axon diameters, unless accounted for [24 28 29].

The purpose of this study was to determine the features of non-straight axons that are observable with dMRI and explore how these may confuse an analysis based on the straight-cylinder assumption. Our investigation utilized a toy-model of axons and their trajectories, which we refer to as undulating thin-fiber model, that enabled us to study effects of the non-straight trajectories on the time-dependence of intra-axonal diffusion. The time-dependence was characterized using the frequency analysis of signal attenuation in the first order cumulant expansion, where signal attenuation is determined by the inner product of the encoding spectrum and the diffusion spectrum [9]. Our analysis focused on four topics. First, we used the undulating thin-fiber model to investigate theoretically which properties of the trajectories determine the characteristics of the diffusion spectrum. Second, we compared results from the theoretical approach to those from numerical simulations. Third, we compared the diffusion spectra from our fiber model to the ones arising from cylinders [8] and from the short-range disorder [17 18], representing intra- and extra-axonal diffusion, respectively. Fourth, we assessed when and why the models that assume straight cylinders overestimate the axon diameter in the presence of axonal undulations.

## Theory

We represent axons by a simple toy-model comprising spin-carrying, infinitesimally thin, and infinitely long undulating fibers. Let the undulating thin-fiber be given by the longitudinal ($x$) and transversal ($y$) coordinates, related by

$$y(x \mid a, \lambda, \varphi) = a \cdot \sin\left(2\pi \cdot \frac{x}{\lambda} + \varphi(x)\right), \qquad (1)$$

where $a$ is the undulation amplitude, $\lambda$ is the undulation wavelength, and $\varphi(x)$ is a phase-modulating factor representing stochastic variations. The toy-model was inspired by histology of brain tissue, which shows nearly sinusoidal axonal trajectories with undulation amplitudes an order of magnitude larger than the axon diameter (Figure 1).

We used the undulating thin-fiber model to analyze three cases with increasing complexity, using both theoretical tools and numerical simulations. The first case comprised fiber trajectories described by a single amplitude and wavelength and a constant phase-modulating factor (1-harmonic case). The second case comprised multiple 1-harmonic fiber trajectories characterized by a distribution of amplitudes and wavelengths, again with a constant phase (n-harmonic case). The third case comprised 1-harmonic trajectories with a stochastic phase (the stochastic case).

**Diffusion spectrum**

Properties of the diffusion spectrum was analyzed for all three cases in the direction transversal to the main fiber direction, because it is this component that contributes to dMRI-based estimates of the axon diameter [30]. The signal attenuation can then be analyzed in terms of two factors [9]: one that describes the diffusion in the system (the diffusion spectrum) and one that determines which parts of the diffusion spectrum are encoded into the signal (the encoding power spectrum, also known as the dephasing power spectrum). The diffusion spectrum is defined as the Fourier transform of the velocity autocorrelation function [9]:

$$D(f) = \mathcal{F}\{\langle v(t)v(0)\rangle\} = \frac{1}{2}\int_{-\infty}^{\infty} \langle v(t)v(0)\rangle \cdot e^{-2i\pi \cdot t \cdot f}\, dt. \qquad (2)$$

The velocity autocorrelation function $\langle v(t)v(0)\rangle$ is in turn related to the mean square displacement $\langle \Delta y^2(t)\rangle$ as

$$\langle v(t)v(0)\rangle = \frac{1}{2}\frac{d^2}{dt^2}\langle \Delta y^2(t)\rangle. \tag{3}$$

The quantities – the mean square displacement $\langle \Delta y^2(t)\rangle$, the velocity autocorrelation function $\langle v(t)v(0)\rangle$, the diffusion spectrum $D(f)$ and also time-dependent diffusion coefficient $D(t)$ are related and provide the same information, but the diffusion spectrum is most directly related to the signal attenuation. Up to the first order (that is, up to moderate attenuations), the attenuation is given by the inner product of the diffusion spectrum $D(f)$ and the encoding power spectrum $|q(f)|^2$

$$S \approx \exp\left(-\int_{-\infty}^{+\infty} D(f) \cdot |q(f)|^2 \, df\right). \tag{4}$$

The encoding spectrum $q(f)$ is defined from a gradient waveform $g(t)$ as

$$q(f) = \mathcal{F}\{q(t)\} = \mathcal{F}\left\{\gamma \int_0^\tau g(t)\, dt\right\} = \int_{-\infty}^{+\infty} q(t)\, e^{-2\pi i \cdot t \cdot f}\, dt \tag{5}$$

where $\tau$ is the echo-time. For completeness, note that the b-value is the total encoding power given by

$$b = \int_{-\infty}^{+\infty} |q(f)|^2 \, df. \tag{6}$$

**Characterizing diffusion spectrum**

Our first aim was to predict features of the diffusion spectra from the parameters of the undulating thin-fiber model. Corresponding features of the spectrum for a cylinder were used for reference. Four spectral features were considered: the spectral width ($f_\Delta$), the spectral height ($D_{hi}$), the simplified spectral shape, and the low-frequency behavior in terms of the power law exponent ($p$). An additional feature was also considered: the diffusivity at zero frequency (infinite diffusion times). However, this was trivially zero because the maximal mean square displacement is bounded by the outermost positions of the fiber trajectories in our fiber model. In the presence of (macroscopic) orientation dispersion of the main fiber directions of different fibers, this would not hold true, but we limited our analysis to the case without macroscopic orientation dispersion.

*Spectral height*

The high frequency limit $D_{\text{hi}}$ represents diffusivity at short times (high frequencies), where the diffusion is unrestricted. For cylinders, $D_{\text{hi}}$ is trivially given by the bulk diffusivity $D_0$. For all cases of our fiber model, $D_{\text{hi}}$ can be obtained by noting that in the limit of short diffusion times we can approximate an undulating thin-fiber by straight but orientationally dispersed and disconnected segments. The diffusion coefficient perpendicular to an angulated segment is given by

$$D(t \to 0) = \langle \sin^2(\theta(x)) \rangle \cdot D_0 \qquad (7)$$

where $x$ is the position of the segment, $\langle \cdot \rangle$ denotes averaging over all segments, $\theta$ is the angle between the direction of the segment and the main fiber direction. We define the microscopic orientation dispersion µOD as

$$\mu\text{OD} = \langle \sin^2(\theta(x)) \rangle . \qquad (8)$$

Examples of µOD values are shown in Table 2. For all three cases, we can thus predict that

$$D_{\text{hi}} = \mu\text{OD} \cdot D_0 \qquad (9)$$

Note that this would imply that $D_{\text{hi}} < D_0$ also for infinite frequencies (e.g. the diffusion is never free), which is absurd but shows the limit of our toy-model. That is why we denote the spectral height by the symbol $D_{\text{hi}}$ and not by e.g. $D_\infty$. The asymptotic behavior limits the scope of our fiber model, which will we will bring up again in the discussion.

It is also worth noting that for the n-harmonic case, $D_{\text{hi}}$ is given by the average of contributions from $n$ different 1-harmonic fibers

$$D_{\text{hi}} = \langle D_{\text{hi};\,i} \rangle \approx D_0 \cdot \langle \mu\text{OD}_i \rangle \qquad (10)$$

*Spectral width*

The spectral width captures the frequency where the time-dependent diffusion effects have the highest impact on the signal. For cylinders, the spectral width ($f_\Delta$) is inversely related to the time required for the mean square displacement ($\langle \Delta y^2(t) \rangle = 2D_0 t$) to approach the fiber diameter ($d$)

$$f_\Delta = k_\text{c} \cdot \frac{D_0}{d^2} \,[\text{Hz}] \qquad (11)$$

where $k_c = \sqrt{1536/7 \cdot 1/4\pi^2} \approx 2.35$ is a proportionality constant determined by analyzing the infinite sum of the analytical diffusion spectrum of cylinders [9 17]. The spectral width is thus an important parameter in the cylinder case because it is directly related to the cylinder diameter.

For the undulating thin-fiber model, we assess each case separately. For the 1-harmonic case, the diffusion process is similar to that in the cylinder because the outer limits of the fiber trajectory will act as reflecting boundaries. However, longer diffusion times are needed to reach the boundaries as the water molecules cannot diffuse along the shortest paths but are bounded by the fiber trajectory. We thus postulate that the spectral width is

$$f_\Delta \approx k_h \cdot \frac{\widetilde{D}_0}{a^2} \tag{12}$$

where $k_h$ is a proportionality constant, $a$ is the undulation amplitude and $\widetilde{D}_0$ is the apparent bulk diffusivity coefficient. We did not find an analytical expression for $k_h$, but determined it from the numerical simulations. The bulk diffusivity coefficient $D_0$ is reduced by the path length ratio of the straight path and the one dictated by the fiber

$$\widetilde{D}_0 = D_0 \cdot \frac{a^2}{\left(\int_C y(x)\,dx\right)^2} \tag{13}$$

where the denominator denotes a path integral over one wavelength of the sine wave. The path length ratio can be linked to microscopic orientation dispersion µOD

$$\frac{a^2}{\left(\int_C y(x)\,dx\right)^2} = \frac{\langle |dy(x)|\rangle^2}{\langle dl\rangle^2} = \langle \frac{dy^2(x)}{dl^2}\rangle = \langle \sin^2(\theta(x))\rangle = \mu OD \tag{14}$$

where $dl$ is the length of a segment ($dl^2 = dx^2(x) + dy^2(x)$) that is in our case kept constant. Taken together, we predict the spectral width for the 1-harmonic case is given by

$$f_\Delta \approx k_h \cdot \frac{D_0}{a^2} \cdot \mu OD \tag{15}$$

For the n-harmonic case, we first note that fibers with higher $D_{hi}$ have a larger impact on the spectral width $f_\Delta$. The width can thus be approximated by a weighted average of the 1-harmonic spectral widths $f_{\Delta;i}$, where the weight $w_i$ is given by the spectral height $D_{hi;i}$, according to

$$f_\Delta \approx \frac{\langle w_i \cdot f_{\Delta;\,i}\rangle}{\langle w\rangle} = \frac{\langle D_{\text{hi};\,i} \cdot f_{\Delta;\,i}\rangle}{\langle D_{\text{hi};\,i}\rangle} = k_\text{h} \cdot D_0 \cdot \frac{\langle \mu OD_i^2/a_i^2\rangle}{\langle \mu OD_i\rangle}. \tag{16}$$

The stochastic case can be analyzed similarly to the n-harmonic case (Eq. 16), by assuming each segment of the stochastically undulating fiber to have its own spectral height and width. These were assumed to be given by Eqs. 9 and 15, but using the local microscopic orientation dispersion and amplitude that is given by the maximal deviation from the straight path ($a_\text{max}$). The local microscopic orientation dispersion was defined as $\mu OD(x) = \sin^2(\theta(x))$. The spectral width can then be approximated by averaging all local segments using the same approach as in Eq. 16, as

$$f_\Delta \approx k_\text{s} \cdot \frac{\widetilde{D}_0}{a_\text{max}^2} = k_\text{s} \cdot \frac{D_0}{a_\text{max}^2} \cdot \frac{\langle \mu OD^2(x)\rangle}{\mu OD} \tag{17}$$

where $k_\text{s}$ is a proportionality constant that can be determined from simulations and where $\widetilde{D}_0$ is given by

$$\widetilde{D}_0 = D_0 \cdot \frac{\langle \mu OD^2(x)\rangle}{\langle \mu OD(x)\rangle} \tag{18}$$

where $\langle \mu OD(x)\rangle = \mu OD$.

*Simplified spectral shape and the low-frequency behavior*

To analyze the shape of the diffusion spectrum, we represent it by an approximation and study the differences between the simplified and simulated spectra. For simple geometries such as parallel planes, cylinders and spheres, the diffusion spectra are given analytically by an infinite sum of Lorentzians [9]

$$D(f) = \sum_{k=1}^{\infty} D_{\text{hi};\,k} \cdot \frac{f^2}{f_{\Delta;\,k}^2 + f^2} = D_0 \sum_{k=1}^{\infty} a_k B_k \cdot \frac{f^2}{(a_k D_0)^2 + f^2} \tag{19}$$

where $a_k$ and $B_k$ are coefficients related to the geometry

$$a_k = \left(\frac{\zeta_k}{r}\right)^2 \qquad B_k = \frac{2(r/\zeta_k)}{\zeta_k^2 + 1 - \text{dim}} \tag{20}$$

and where $2 \cdot r$ is the diameter and $\zeta_k$ are the kernels of

$$\zeta J_{\text{dim}/2-1}(\zeta) - (\text{dim} - 1) J_{\text{dim}/2}(\zeta) = 0 \tag{21}$$

where $J_\upsilon$ denotes the $\upsilon$-order Bessel function of the first kind, $\text{dim} = 1, 2$ or $3$ for planar, cylindrical or spherical restrictions, respectively. Although Eq. 19 contains an infinite sum, we note that the majority of the spectrum is accounted for by the first Lorentzian term, given by the coefficient $a_1 B_1$, which is 0.83 for cylinders. A simplified spectrum can thus be defined by a single Lorentzian approximation

$$D(f) \approx L(f) = D_{hi} \cdot \frac{f^2}{f_\Delta^2 + f^2} \quad (22)$$

where $D_{hi}$ and $f_\Delta$ are the two features determining to the spectral height and width, respectively. This simplified spectrum will be used to study dissimilarities between cylinders and the undulating thin-fiber model.

Another aspect of the simplification of the spectra is its low-frequency behavior, which contains information on the gradual coarse-graining process over structural length-scales [31-34]. The single Lorentzian approximation can be expanded to the second-order as

$$L(f) \approx \frac{D_{hi}}{f_\Delta^2} \cdot f^2. \quad (23)$$

Plugging in the results for the cylinder case ($D_{hi} = D_0$ and Eq. 11) yields a proportionality

$$L(f) \approx \frac{1}{k_c^2} \cdot \frac{d^4}{D_0} \cdot f^2 [\text{Hz}] \quad (24)$$

and in the 1-harmonic case (Eqs. 9 and 15)

$$L(f) \approx \frac{1}{k_h^2} \cdot \frac{a^4}{\mu OD \cdot D_0} \cdot f^2 \, [\text{Hz}]. \quad (25)$$

Under the assumption that the single Lorentzian approximation describes well the spectra for the undulating thin-fiber model, the parameters that determine the low-frequency behavior are $a$ and µOD. The diameter thus relates to the parameters in the 1-harmonic case of our fiber model as

$$d \approx \sqrt{\frac{k_c}{k_h}} \cdot \frac{a}{\sqrt[4]{\mu OD}} \quad (26)$$

This relation suggests that undulations due to non-zero microscopic orientation dispersion can bias diameter estimations. Note that Eq. 26 assumes that the low-frequency second-order expansion (Eq. 23) is an adequate simplification of the diffusion spectrum in the range of frequencies attainable in typical dMRI experiments, which is not guaranteed. We have thus also investigated the low-frequency behavior by also using simulations.

## Methods

Numerical simulations were performed to test the theoretical predictions. Diffusion spectra were generated for each case by first computing mean square displacement $\langle \Delta y^2(t) \rangle$. After that, the velocity autocorrelation function $\langle v(t)v(0) \rangle$ was calculated from the mean square displacement using Eq. 3. Finally, $\langle v(t)v(0) \rangle$ was Fourier-transformed (Eq. 2) to obtain the diffusion spectrum $D(f)$. The numerical analysis was implemented in Matlab (The Mathworks, Natick, USA) and is available at https://github.com/jan-brabec/undulating_structures.

### Substrate definition

The parameters of the undulating thin-fiber model in Eq. 1 were selected to represent a range of biologically relevant conditions. The parameters were obtained from histological images of corpus callosum (Table 1). However, we acknowledge that the numbers should be considered approximate, since axonal trajectories have not yet been systematically investigated from a histological perspective.

The pathways defined by our fiber model were discretized into straight segments with lengths of $dl = 0.1$ μm. For the 1-harmonic case, we investigated undulation amplitudes of $a = 1, 2$ and $3$ μm and wavelengths of $\lambda = 10, 20, 30, 40$ and $50$ μm. For the n-harmonic case, we assumed that the 1-harmonic fibers are well simplified by the single Lorentzian approximation (Eq. 22), which we will brought up in the result section. The amplitudes and wavelengths were drawn from gamma distributions with $n = 1000$. Here, the shape (α) and scale (β) parameters of two gamma distributions were drawn from uniform distributions: $\alpha_a \sim U(0, 10)$, $\alpha_\lambda \sim U(0, 10)$, $\beta_a \sim U(0, 3 \cdot 10^{-6})$ and $\beta_\lambda \sim U(0, 50 \cdot 10^{-6})$. Second, the amplitude and wavelength distributions were formed as the gamma distributions

$$a \sim \Gamma(\alpha_a, \beta_a) + 1 \text{ μm,} \tag{27}$$

$$\lambda \sim \Gamma(\alpha_\lambda, \beta_\lambda) + 10 \text{ μm.} \tag{28}$$

Both samples were then restricted to the same parameters as investigated for the 1-harmonic case

$$a \rightarrow a(a > 1 \text{ μm} \,\&\, a < 3 \text{ μm}), \tag{29}$$

$$\lambda \rightarrow \lambda(\lambda > 10 \text{ μm} \,\&\, \lambda < 50 \text{ μm}). \tag{30}$$

For the stochastic case, variations were introduced into the harmonic sine waves by modeling the phase-modulating factor $\varphi(x)$ in Eq. 1 as a first-order autoregressive process AR(1): $\varphi(x_i) = \rho \cdot x_{i-1} + \varsigma(x)$, where $\varsigma(x)$ are independent normally distributed random numbers. The randomly generated numbers were normalized and cumulatively summed. To avoid high frequency fluctuations that would represent unphysical turns of the fiber, the random numbers were smoothed using a moving average filter with a width of 0.5 µm.

**Numerical simulations**

*Gaussian Sampling*

To estimate the mean square displacement effectively, we implemented what we refer to as the Gaussian sampling method. It assumes one-dimensional Gaussian diffusion along the fiber trajectory, so that the displacements are described with a normal distribution with a mean of zero and a variance of $\sigma^2 = 2D_0 t$, with the bulk diffusion coefficient set to $D_0 = 1.7$ µm²/ms. The mean squared displacements perpendicular to the main fiber direction $\langle \Delta y^2(t) \rangle$ were then computed by the following pseudo-algorithm:

- for $t = 0 : dt : t_{max}$
    - $\sigma = (2D_0 t)^{1/2}$
    - for $l = l_{min} : dl : l_{max}$
        - $\Delta l = -4 \cdot \sigma : dd : 4 \cdot \sigma$
        - $\Delta y = y(l + \Delta l \mapsto x) - y(l \mapsto x)$ (according to Eq. 1)
        - $w = \exp(-\Delta l^2 / 2\sigma^2)$
        - $\langle \Delta y^2(t) \rangle = \langle \Delta y^2(t) \rangle + dl/(l_{max} - l_{min}) \cdot \sum w \cdot \Delta y^2 / \sum w$.

Bold italic letters denote vectors, multiplication of vectors is element-wise, "$\mapsto$" represents mapping from one coordinate system to other and $\langle \Delta y^2(t) \rangle$ is a symbol representing the output variable. The procedure was repeated in discrete time steps with $dt = 100$ µs for times up to $t_{max} = 1$ s in the 1-harmonic case. In the stochastic case, we used $t_{max} = 10$ s due to slower convergence. The parameter $dd$ was set to $dd = dl = 0.1$ µm. In the frequency domain, these numbers correspond to a maximal frequency of $f_{max} = 5000$ Hz with $df = 1$ Hz for $t_{max} = 1$ s and to $df = 0.1$ Hz for the case with $t_{max} = 10$ s. The computation was initiated within one period of the sine wave for the 1-harmonic and n-harmonic cases ($l_{min}$ to $l_{max}$). In the stochastic case, a representative sample of the whole structure must be probed for the end result to be representative (ergodic condition). An additional simulation was used to find that the sum of diffusion spectra contributions in Eq. 2 converged within 30 wavelengths of the underlying harmonic sine waves ($l_{min}$ to $l_{max}$).

**Validation of diffusion spectra and signal simulations**

Our Gaussian sampling approach was verified against an alternative approach using Monte Carlo simulations, in which particles were simulated as exhibiting free diffusion along a one-dimensional fiber. The positions along the fiber were then remapped to $x$ and $y$ coordinates using Eq. 1 and used to calculate mean square displacement as a function of time. The same fiber trajectories as used with the Gaussian sampling method were used in the Monte Carlo simulator, with the same discretization parameters, except for the higher temporal resolution, $dt = 10$ µs, corresponding to $f_{max} = 50$ kHz in the frequency domain. The Monte Carlo simulations were performed with $10^6$ particles.

The forward model of the signal was also verified against Monte Carlo simulations. The signal was generated from synthetic diffusion spectra based on the semi-analytical Gaussian sampling method (Eq. 4). This signal was then compared with the signal obtained from Monte Carlo simulations by accumulating the phase ($\Phi$) according to

$$\Phi(t) = \int_0^t g(t')x(t')dt' \tag{31}$$

yielding attenuation as

$$S = \langle \exp(-i \cdot \Phi) \rangle. \tag{32}$$

Gradient waveforms $g(t)$ for the protocols by Alexander et al. [8] were used (see Table 3).

**Characterization of the diffusion spectrum**

Four features of the diffusion spectra were characterized and compared: spectral height, width, simplified spectral shape, and low-frequency behavior.

*Spectral height and width*

Spectral heights and widths were predicted from the parameters of our fiber model by Eqs. 9, 15 in the 1-harmonic case, Eqs. 10, 16 in the n-harmonic and Eqs. 9, 17 in the stochastic case, respectively. These theoretical predictions were compared with the corresponding values estimated from the simulated spectra. Spectral widths for the 1-harmonic (Eq. 15) and stochastic (Eq. 17) cases were first predicted assuming $k_h = k_s = 1$, the obtained data was fitted with a first degree polynomial and adjusted by the obtained proportionality constants. Spectral heights were estimated from the simulated diffusion spectra as the average of D($f$) within 900 Hz to 1000 Hz. Within this frequency range, the spectra were fitted by a first-degree polynomial to ensure that they reached stationary value (slope less than $10^{-13}$). Spectral widths were estimated as the closest frequency corresponding to the half of the spectral height

$$D(f_\Delta) = \frac{1}{2} D_{hi} \tag{33}$$

*Simplified spectral shape*

To test whether there are any relevant differences in the shape of the diffusion spectra between simulated and simplified spectra by the single Lorentzian approximation (Eq. 22), we investigated their difference in predicted signal values and compared this difference to that expected from pure noise. Because the simplified spectrum has only two free parameters ($D_{hi}, f_\Delta$), the spectra may differ in a way that is not captured by these two parameters. To limit ourselves to the differences that could potentially be detected on a typical clinical scanner, we investigated the differences from the single Lorentzian approximation with SNR = 50. Four simulation setups were investigated: undulating thin-fibers from the 1-harmonic case ($a = 2$ μm, $\lambda = 30$ μm), the n-harmonic case and the stochastic case, as well as cylinders with diameter $d = 20$ μm. All cases were chosen so that they had a similar spectral width and height (1-harmonic: 10 Hz and 0.13 μm²/ms, n-harmonic: 8 Hz and 0.14 μm²/ms, stochastic: 7 Hz and 0.13 μm²/ms and cylinders: 11 Hz but 1.7 μm²/ms). We then computed the mean squared error (MSE) between the signals obtained from simulated and simplified diffusion spectra, defined as

$$\text{MSE} = \langle (S_{L(f)} - S_{D(f)})^2 \rangle \tag{34}$$

Note that due to normally-distributed noise $\varepsilon$ at given SNR = 50 we have

$$\varepsilon \sim N(0, \sigma^2); \ E(\varepsilon^2) - E(\varepsilon)^2 = V(\varepsilon) - 0 = V(\varepsilon) = \sigma^2 \tag{35}$$

$$\text{MSE} = \langle \varepsilon^2 \rangle = \sigma^2 = 4 \cdot 10^{-4}. \tag{36}$$

If the MSE is below $4 \cdot 10^{-4}$, the difference in predicted signals are indistinguishable from noise, for the protocol in question.

*Low-frequency behavior*

Although we studied the experimental detectability of differences that include also the low-frequency region, we focused on the properties in this region in particular because it may reveal valuable structural information regardless whether the properties are detectable by our protocol [31 33 34]. However, we studied the low-frequency behavior from an experimentalist's point of view. We estimated the exponents $p$ by fitting because we intended to use similar methodology and obtain results that can be compared with other works in the field (Burcaw et al. [17]). To assess the robustness of our estimation of the exponent $p$, we used two different frequency regions for fitting. Also, if the spectral widths were small, we lowered the maximum frequency to a fraction of the spectral height where the exponents $p$ was fitted to consider that the exponent should reflect, indeed, the behavior of the *low*-frequency part of the spectra. Specifically, we compared the low-frequency region of the simulated diffusion spectra between the cases of the undulating thin-fiber model with cylinders, where the spectra can be approximated as

$$D(f) \approx f^p, \qquad (37)$$

where $p > 0$ is a real number. The power law exponent $p$ was fitted in two regions: at the *very low*-frequency region: $[0 \text{ Hz}; \min(0.2 \cdot D_{\text{hi}}; 5 \text{ Hz})]$ and *low*-frequency region: $[0 \text{ Hz}; \min(0.5 \cdot D_{\text{hi}}; 20 \text{ Hz})]$.

**Implications for axon diameter mapping**

We studied the link between the parameters of the undulating thin-fiber model and cylinder diameters estimated by the ActiveAx model. Simulated diffusion spectra of the 1-harmonic case were used to generate signal data based on the Alexander protocol [8], assuming absence of noise (infinite signal-to-noise ratio). The ActiveAx model was constrained to the case of intra-axonal diffusion only and the diameters were estimated by fitting the simulated signal. The estimated diameters were plotted against parameters characterizing our fiber model: undulation amplitude $a$ and Eq. 26 relating the diameter $d$, amplitude $a$ and microscopic orientation dispersion based on a second-order Taylor expansion.

## Results

### Validation of the Gaussian sampling method

Comparisons between the proposed Gaussian sampling method and regular Monte Carlo simulations showed a high agreement between the methods, but the newly proposed method offered a superior computational speed (2 days for Monte Carlo compared with 2 minutes for Gaussian Sampling for the same discretization and fiber settings). The simulated diffusion spectra for the two methods overlapped (Figure 2A), although the ones obtained by Monte Carlo simulations showed more noise, despite using one million particles in the simulation and ten times higher temporal resolution. The two methods also showed high agreement for the simulated signal attenuations (Figure 2B), even though quite different methods were used to compute the signal (via the diffusion spectrum in Eq. 4 for the Gaussian method and via simulated phase distributions with Eq. 32 for the Monte Carlo method). For the protocol optimized in [8] and employed in this study, three of the gradient waveforms have encoding widths of 6 Hz and one of 20 Hz. They sensitize the signal up to frequencies of only 10 Hz and the remaining one up to 50 Hz (Figure 2C, black and purple curve, respectively). Finally, we note that the Gaussian approximation of the cumulant expansion (Eq. 4) yield in the worst case identical results as the Monte Carlo simulations for signal attenuations up to 60 % (Figure 2D) [35]. This worst-case example means that the spectral approach is valid within practically achievable b-values up to 10 ms/µm$^2$.

### Characterizing diffusion spectrum

Figure 3 shows spectral heights estimated from Gaussian sampling simulations versus the theoretically predicted values (Eqs. 9 and 10). The estimated and predicted spectral heights showed a high agreement, which verified that the spectral height is determined by the microscopic orientation dispersion. Figure 4 shows the corresponding analysis for the spectral width, with the theoretical predictions according to Eq. 15 for the 1-harmonic case, Eq. 16 for the n-harmonic case, and Eq. 17 for the stochastic case of our fiber model. Proportionality constants were found to be $k_\mathrm{h} \approx 0.34$ for the 1-harmonic and n-harmonic cases and $k_\mathrm{s} \approx 0.13$ for the stochastic case. Estimated values agreed with theoretical predictions after adjustment with this proportionality constant, however, the residual variance was larger for the stochastic case compared to the other cases.

Figure 5 shows diffusion spectra for a cylinder and for the 1-harmonic, n-harmonic and the stochastic cases of our fiber model. These examples were selected to have similar spectral heights and widths to highlight differences apart from the widths and heights. Despite some differences at low and high frequencies, all of the simulated spectra were well simplified by the single Lorentzian approximation $L(f)$ (Eq. 22), at least qualitatively. Quantitatively, the differences in signals generated from simulated or simplified spectra were highly similar with a difference indistinguishable from that induced by noise, for SNR $\leq$ 50 (i.e. MSE $< 4 \cdot 10^{-4}$ smaller than noise in all cases; Table 4) with 1-harmonic case being the closest to the simplified spectrum. This means that the studied examples, the differences between the simulated and simplified spectra cannot be experimentally distinguished under realistic noise levels when using the protocol in Table 3.

Figure 6 highlights differences in the low-frequency regime between the spectra for the 1-harmonic, n-harmonic, and stochastic cases. In Figure 6A, we see that the n-harmonic and stochastic spectra approach the spectral height slower than the 1-harmonic spectra. The low-frequency behavior in the 1-harmonic case generally resembles the low-frequency behavior in the case of cylinders with power law exponent $p = 2$. The stochastic and n-harmonic cases, however, have $p < 2$ (Figure 6B). The exact value of the exponent $p$ depends, however, on the region in which the slope was evaluated. Fitting the diffusion spectra in the frequency range up to 5 Hz yielded a value of the exponent $p$ close to 2 for the 1-harmonic case, but below 2 for the n-harmonic and stochastic cases (Figure 6C). Fitting in the frequency range up to 20 Hz yielded substantially lower values (Figure 6D).

Figure 7 illustrates the dependence of the exponent $p$ in the n-harmonic case on the distribution of undulation amplitudes and wavelengths. Results show that if the n-harmonic fibers are formed from a narrow distribution of amplitudes and wavelengths, they resemble a 1-harmonic spectrum (Figure 7A), but that a dispersed distribution yields a reduced exponent $p$ (Figure 7B). When nearly straight fibers are introduced (characterized by $a < 1$ μm and $\lambda > 50$ μm, adjusting the steps described by the Eqs. 29 and 30), the resulting spectrum resembles the one associated with the stochastic case (Figure 7C).

**Implications for axon diameter mapping**

Figure 8 shows cylinder diameters estimated by a model that assumes straight cylinders from signal data for the 1-harmonic case of the undulating thin-fiber model. Expected values of the diameter are zero in the undulating thin-fibers, but the estimated diameters are substantially higher even in the presence of slight deviations from straight paths (ratio between undulation amplitude and wavelength between 2 and 30 %). Estimated diameters correlated with the undulation amplitude (Figure 8A). The relation is not linear with respect to the wavelength $\lambda$ because the overestimation is weakest for 10 µm and strongest for wavelength $\lambda = 30$ µm but not for $\lambda = 50$ µm. Hypothetically, a link between parameters of our fiber model and cylinder diameters can be established using a second-order Taylor expansion (Eq. 26). However, the hypothetical relationship holds only in some cases while in others the diameter is systematically underestimated (Figure 8B). Figures 8C and 8D show the encoding spectra as well as the diffusion spectra for the cylinder and 1-harmonic case of our fiber model, with the difference that the hypothetical relationship was accurate for the undulating thin-fiber used in Figure 8C (star marker in Figure 8B) but not for 8D (square marker in Figure 8B). Comparing encoding and diffusion spectra shows why. In Figure 8C, all of the encoding power is within the quadratic part of the diffusion spectrum, where the Taylor expansion is accurate. In Figure 8D, some of the encoding power is found at the part of the spectrum where the Taylor expansion is no longer valid.

## Discussion

Undulating thin-fibers have a diffusion-time dependence that is similar to that in cylinders, at least for lower frequencies (long diffusion times). This can lead to overestimated axon diameters when interpreting data from undulating axons using models that represent axons by straight cylinders, and this positive bias is non-trivially associated with the undulation parameters (Figure 8). Observed effects of time-dependent diffusion in brain white matter are subtle [36], and may potentially be attributed to undulation parameters rather than an axon diameter. For example, axon diameter indices of 3–12 micrometers were found by application of the ActiveAx model to the corpus callosum in human and monkey brains [8]. Similar results can be explained solely by the presence of weak undulations (Figure 8) of axons with diameters below the resolution limit of 3-5 microns [19], which is a biologically plausible case. The reason why undulations can be misinterpreted as axon diameters is that undulating thin-fibers and cylinders have similar diffusion spectra in the region sensitized by popular diffusion encoding protocols (Figure 2, Table 3). Previous similar work by Nilsson et al. [24] studied the effects of axonal undulations on the diffusion propagator, and found the width of the propagator to reflect the undulation amplitude rather than the cylinder diameter. Here we studied the effects of undulations from a broader perspective and found the microscopic orientation dispersion of the thin-fibers to be the most useful toy-model parameter to predict features of the diffusion spectra. Our results were based on the frequency rather than the propagator formalism, because the former allows for simpler generalization to any gradient waveform without having to assume the narrow pulses (i.e. $\delta \to 0$) [37]. We also demonstrated the overestimation by using a popular encoding protocol from Alexander et al. [8] (Table 3).

The parameters of the undulating thin-fiber model also affected the low-frequency behavior of the diffusion spectrum. Such features were recently highlighted as important for dMRI, when perspectives from condensed matter physics were introduced into the field. This enabled different types of media to be grouped into a relatively few distinct structural universality classes, defined trough their low-frequency behavior. Specifically, the power law exponent $p$ was shown to capture long-range spatial correlations of restrictions [33]. Burcaw et al. associated the structure of the extra-axonal space with short-range disorder characterized by $p = 1$, whereas the restricted diffusion in the intra-axonal space has $p = 2$ [17 18 31], as also seen in Eq. 24. In our work, we estimated the exponent $p$ from the low-frequency part of diffusion spectra, similar to what has been done previously [17]. We found that the structural disorder associated with $p < 2$ may arise from the intra-axonal space alone, provided that axonal trajectories can be sufficiently well represented by the n-harmonic or stochastic cases of the undulating thin-fiber model (Figures 6 and 7). Thus, the time-dependence of intra-axonal diffusion may resemble that of extra-axonal diffusion that is characterized by $p < 2$ and possibly bias models that assume intra-axonal diffusion to have $p = 2$. This should be investigated in future work both theoretically and experimentally.

This work highlights the importance of considering the whole three-dimensional content of the voxel in modeling (i.e. non-straight axonal trajectories as a feature in the often-omitted third dimension) and shows that the time-dependence of the intra-axonal space cannot be fully assessed from thin cross-sections of nerve tissue (i.e. from a two-dimensional cut-plane orthogonal to the axons). For the parameters of our fiber model, the spectral heights and widths span up to two orders of magnitude (in the 1-harmonic case: $f_\Delta = 4 - 92$ Hz, $D_{\text{hi}} = 0.01 - 1$ µm²/ms) for biologically plausible parameters (1-harmonic case: $a = 1 - 3$ µm, $\lambda = 10 - 50$ µm). Detailed histological investigations of axonal trajectories will be crucial for the prediction of the exact frequency ranges where effects of undulations could appear in practice, and thus potentially used to falsify the model proposed herein. Such works are on-going [20 21].

Apart from analyzing the diffusion, we can also analyze the encoding strategy. Here, we can define an encoding width $e_\Delta$ as the half width at half maximum of the encoding power spectrum, at least for experiments that utilize single diffusion encoding [38]. If the width of the encoding spectrum is substantially larger than the width of the diffusion spectrum, i.e. $e_\Delta/f_\Delta \gg 1$, this means that the spectral height is predominantly encoded into the signal. If the opposite case holds true, $e_\Delta/f_\Delta \ll 1$, the signal attenuation will be zero and no information about the model parameters will be encoded. This is analogous to having a cylinder diameter below the resolution limit [19]. In the case where $e_\Delta/f_\Delta \approx 1$, both the spectral width and height of the diffusion spectrum are encoded into the signal. Another way to express these relationships is to consider a characteristic time for the undulating thin-fiber model ($1/f_\Delta$). If the characteristic time of the experiment ($1/e_\Delta$) is much shorter than that of the system, we will probe mainly the microscopic orientation dispersion of the fiber segments, which determines the spectral height. In the other limit, effects of time-dependence due to restriction dominates and the microscopic orientation dispersion is largely smoothed out by the diffusion. We can also consider how microscopic orientation dispersion contributes to the total orientation dispersion. In the limit, when $e_\Delta/f_\Delta \ll 1$, the total orientation dispersion is dominated by the macroscopic orientation dispersion. In the other extreme, when $e_\Delta/f_\Delta \gg 1$, both micro- and macroscopic orientation dispersion will determine the total orientation dispersion. Models that represent axons as sticks (thin but straight fibers) with some macroscopic orientation dispersion of the main axon directions, such as NODDI [13 14], would thus find different values depending on the value of the ratio $e_\Delta/f_\Delta$.

We would like to highlight six limitations of this work. First, we note that the undulating thin-fiber model is relevant only for the small-diameter axons in brain white matter, but not for larger axons. For larger axons, the intra-axonal time-dependence would occur at similar frequencies as where undulation effects appear. Here, we put the threshold between small and large axons at approximately 4-5 microns. This is also where the resolution limit is found for clinical systems [19], so below this limit axons are well represented by infinitesimally thin-fibers. Note that large axons are present only to a limited extent in the brain but are more common in the spine and for nerves outside the central nervous system. Small axons are found in the corpus callosum [1 39-41], optic nerve [42 43] or phrenic nerve [25 44] (Table 1). The second limitation concerns the overestimated spectral width in the 1-harmonic case for fibers with undulation amplitudes that are large compared to the undulation wavelength (a/λ ≥ 30 %). This limitation may be relevant in extra-cranial nerves and deserves further studies. The third limitation is that the properties of the toy-model were investigated only in the direction perpendicular to the main fiber direction. Preliminary investigations in the parallel direction did, however, not show us any interesting results. The fourth limitation is that macroscopic orientation dispersion of axons was not considered. None of the claims necessitated the presence of macroscopic orientation dispersion, however. Nevertheless, note that the model can be applied in-vivo only when effects of macroscopic orientation dispersion can be accounted for. The fifth limitation is that for the protocol optimized in [8] and employed in this study, the diffusion spectral content above 50 Hz is not encoded into the signal. Waveforms with high-frequency content would be required to distinguish time-dependent diffusion effects due to cylinder-like structures and undulating thin-fibers, because only in cylinders does the spectral height reach the bulk diffusivity. The sixth limitation is that we represent axons by undulating thin-fibers while omitting a wealth of other microscopic features. However, this limitation does not affect the importance of our main claim: that experimental data risk being misinterpreted when using the cylinder model of axons wherever axonal undulations are present. Further experimental work is needed to judge the magnitude of this risk, which could also be regarded as an opportunity to relate diffusion time-dependence to the properties of axonal trajectories.

## Conclusions

We used both theoretical and numerical tools to characterize the time-dependence of diffusion in a toy-model of axons, referred to as undulating thin-fiber model, by quantification of features of the diffusion spectrum. Unfortunately, these features did not represent simple geometrical descriptors of the fiber trajectories, such as the undulation amplitude or wavelength. Rather, the link between the diffusion spectra and the underlying geometries was better understood when a quantity that is important from the point of view of diffusion – microscopic orientation dispersion of the thin-fibers – was considered. Importantly, we found that the characteristics of the diffusion spectra for the undulating thin-fiber model overlapped with those for straight cylinders, at least for the range of lower frequencies that can be probed at most clinical scanners. Without access to data acquired with waveforms having adequate power at high frequencies and without a detailed histological investigation of the axonal trajectories, it is thus challenging to differentiate the effect of undulating axonal trajectories from axon diameters. Moreover, we found that the diffusion spectrum for the undulating thin-fibers could appear similar to that of extra-axon diffusion.

**Acknowledgments**

This research was supported by the Swedish Research Council (grant no. 2016-03443), the Swedish Foundation for Strategic Research (grant no. AM13-0090), Crafoord Foundation (grant no. 20170825), and Random Walk Imaging AB (grant no. MN15).

**Figures**

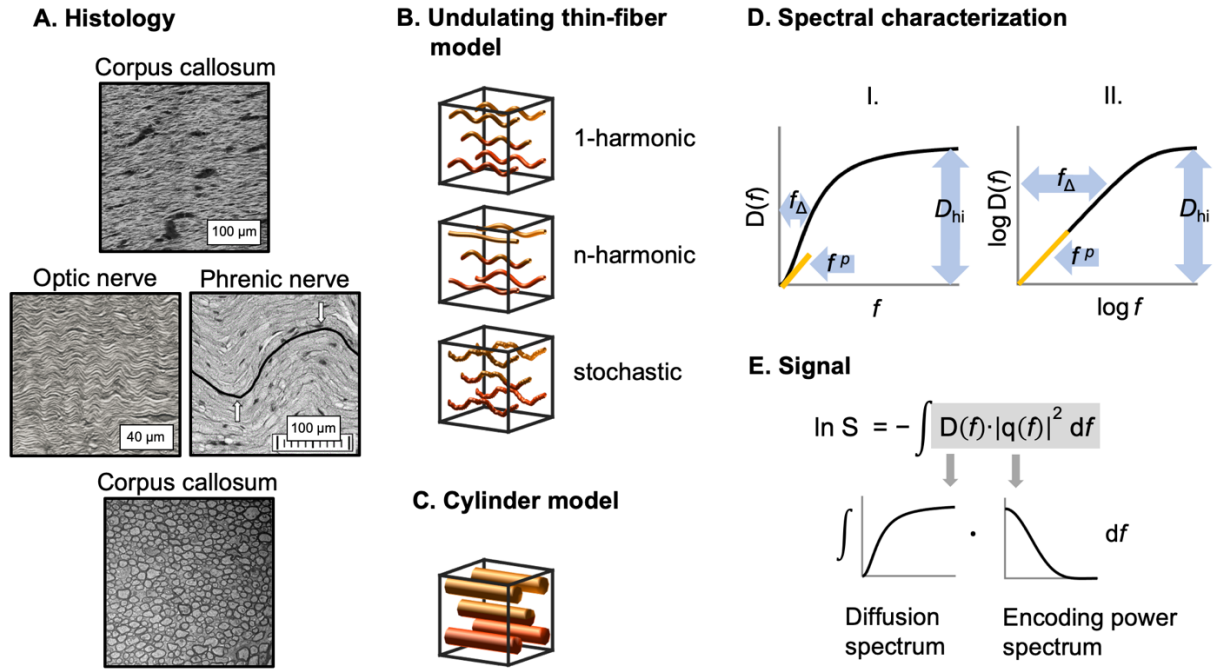

**Figure 1. From histology to diffusion weighted signals.** Panel A shows images of axons, which inspired our undulating thin-fiber model. Axons in the corpus callosum [45], optic nerve [43] and phrenic nerve [25] exhibit ubiquitously sinusoidal undulation patterns. Cross-section of corpus callosum (bottom) [17] implicitly obscures axonal trajectories. Panel B shows three cases of our undulating thin-fiber model: 1-harmonic, n-harmonic and stochastic and comparison to the standard model: non-dispersed straight cylinders in the panel C. Panel D shows the diffusion spectra, which were characterized in terms of their spectral width $f_\Delta$, height $D_{hi}$ and low-frequency behavior ($p$). Panel E shows the connection between diffusion spectra, encoding power spectra and the resulting signal through the first order cumulant expansion. Images in panel A were reproduced with permission from Elsevier, John Wiley & Sons, Inc. and Institute of Electrical and Electronics Engineers (IEEE).

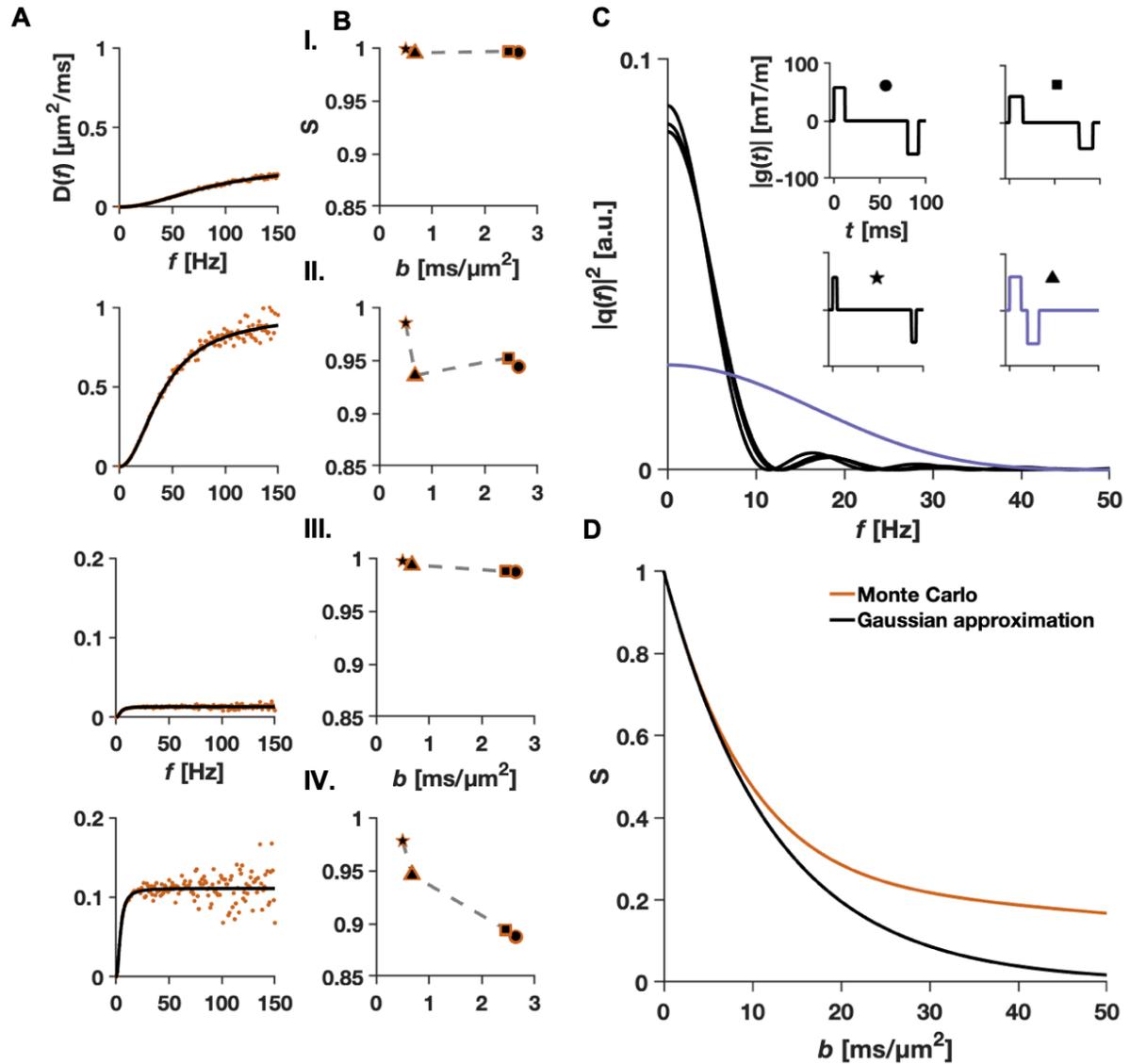

**Figure 2. Comparisons of the proposed Gaussian sampling method and regular Monte Carlo simulations.** Panel A shows diffusion spectra corresponding to the weakest and strongest undulations investigated in the 1-harmonic case (I. $a = 1$ μm, $\lambda = 10$ μm, II: $a = 3$ μm, $\lambda = 10$ μm, III: $a = 1$ μm, $\lambda = 50$ μm and IV: $a = 3$ μm, $\lambda = 50$ μm). The methods showed a high agreement, although the Monte Carlo results (red markers) exhibited more noise and required more computational time than the Gaussian sampling method (black lines). Panel B shows that also the simulated diffusion-weighted signals for the two methods agree. Panel C shows gradient waveforms and their corresponding encoding spectra for the protocol by Alexander et al. [8] (Table 3). The four gradient waveforms probe only two frequency regions [0, 10 Hz] and [0, 50 Hz]. Panel D shows that the first order approximation is in the worst case valid up to attenuations of approximately 60 %.

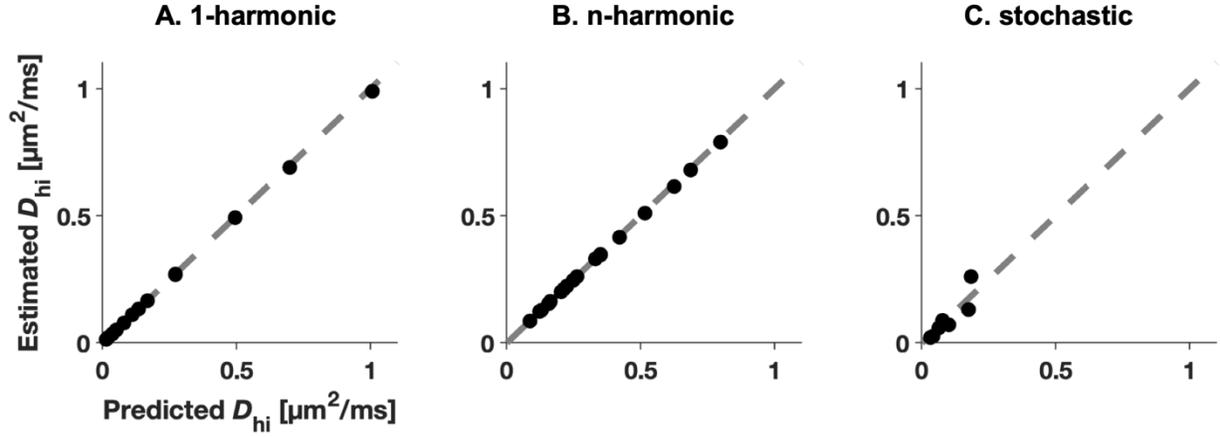

**Figure 3. Predicted and estimated spectral heights $D_{hi}$ are aligned.** Panel A shows the 1-harmonic case, panel B the n-harmonic case and panel C the stochastic case of the undulating thin-fiber model. Values of $D_{hi}$ were predicted using Eq. 9 for the 1-harmonic and stochastic cases and using Eq. 10 for the n-harmonic case.

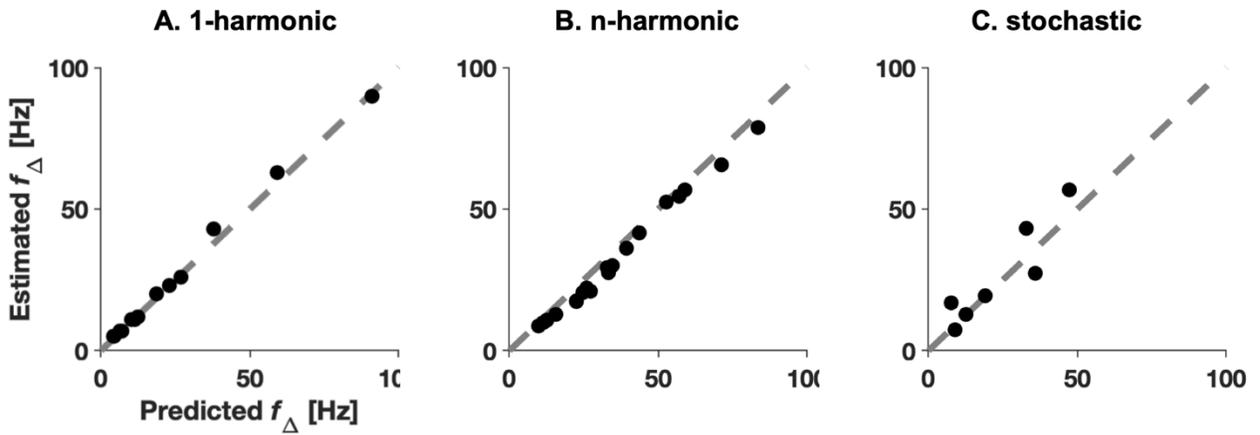

**Figure 4. Predicted and estimated spectral widths $f_\Delta$ are aligned.** Panel A shows the 1-harmonic case, panel B the n-harmonic case and panel C the stochastic case of the undulating thin-fiber model. Values of $f_\Delta$ were predicted using Eq. 15 for the 1-harmonic case, Eq. 16 for the n-harmonic case and Eq. 17 for the stochastic case.

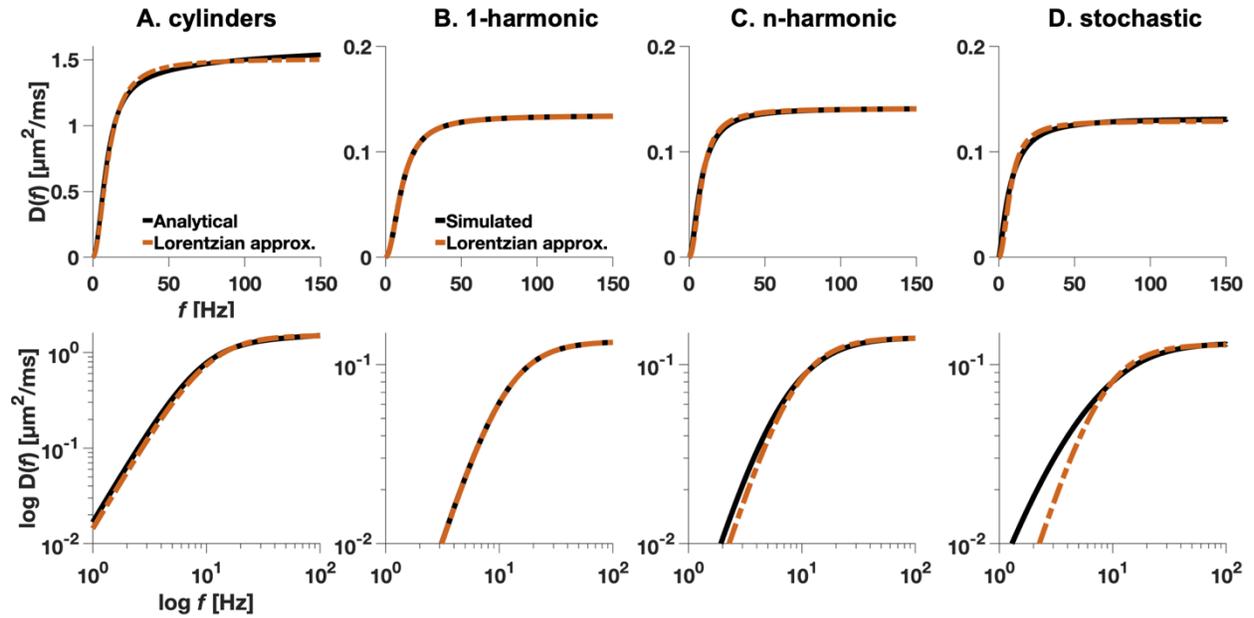

**Figure 5. Diffusion spectra for examples with similar spectral widths and heights for cylinders and all cases of the undulating thin-fiber model compared to the simplified spectra by the single Lorentzian approximation (Eq. 22).** The exact (solid black line) and simplified (dashed red line) diffusion spectra for the case of cylinders (panel A), for the 1-harmonic (panel B), for the n-harmonic case (panel C) and for the stochastic case (panel D) of our fiber model. Bottom row shows the same corresponding spectra as top row but in a log-log plot. The simplified spectra deviate from the simulated ones in the low-frequency range for the n-harmonic and stochastic cases, however, in these examples, the differences are not expected to be detectable within experimental limitations when SNR $\leq$ 50 (Table 4).

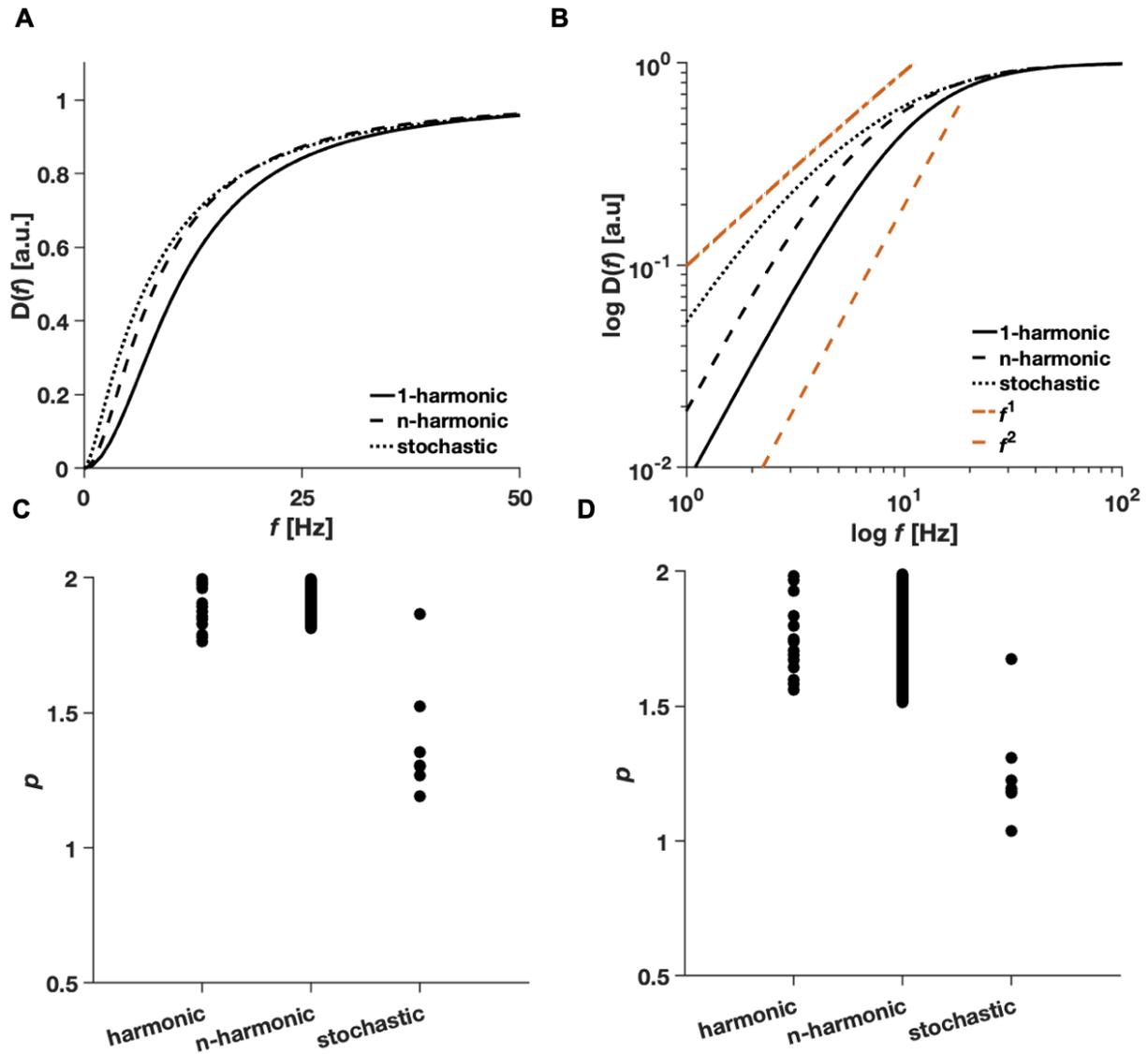

**Figure 6. Diffusion spectra at low-frequencies.** Panel A shows examples of diffusion spectra of the different cases from figure 5 normalized to the same spectral height. Panel B shows the same data but in a log-log plot. The curve of the 1-harmonic case (solid black line) is aligned with the quadratic frequency curve (dashed red line), whereas the slope of the curves from the n-harmonic and stochastic cases (dashed and double dashed black lines) are more aligned with the linear curve (double dashed red line). Panel C shows a distribution of power law exponents $p$ for the simulated cases at the *very low*-frequency region up to 5 Hz. Panel D shows distribution of the exponents $p$ at the *low*-frequency region up to 20 Hz. N-harmonic fibers had gamma distributed undulation amplitudes and wavelengths further restricted to the to the range validated by numerical simulations (1 µm ≤ $a$ ≤ 3 µm; 10 µm ≤ $\lambda$ ≤ 50 µm).

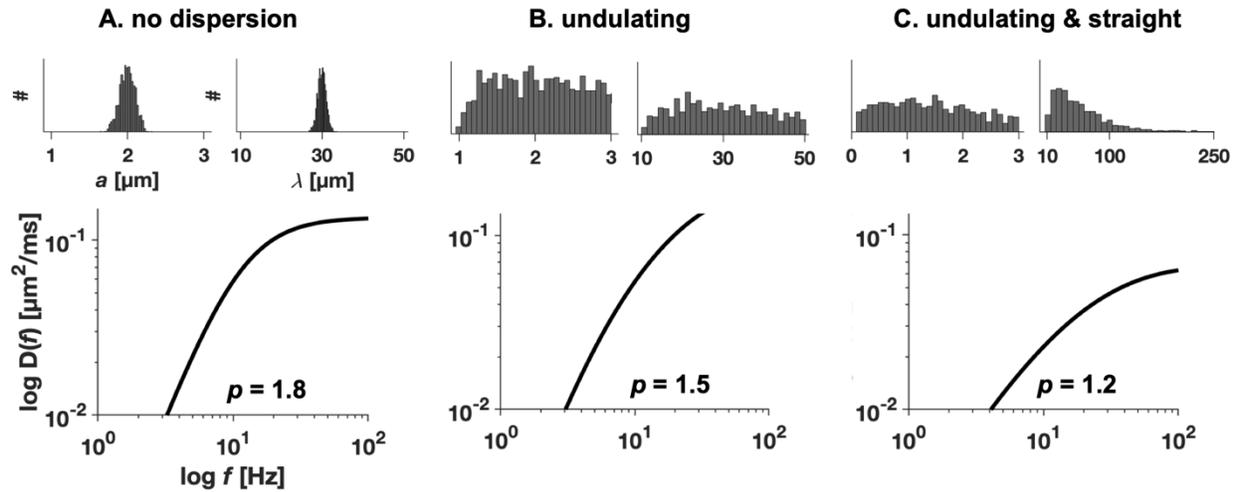

**Figure 7. The low-frequency behavior of the diffusion spectra characterized by the power law exponent $p$ (bottom row) as a function of the distribution of undulation amplitudes and wavelengths for the n-harmonic case (upper row).** Panel A shows that normal distribution of amplitudes and wavelengths with low variance is associated with a spectrum that is similar to the 1-harmonic case. Panel B shows that gamma distribution of amplitudes and wavelengths with parameters within the range validated by numerical simulations yields reduced exponent $p$. This example is also part of the distributions used in Figure 6C and D. Panel C shows that a distribution of amplitudes and wavelengths that contains also almost straight thin-fiber trajectories (i.e. $a < 1$ μm or $\lambda > 50$ μm, adjusting the steps described by the Eqs. 29 and 30) resembles the stochastic spectrum. Exponents $p$ were estimated for spectra at the *low*-frequency region up to 20 Hz.

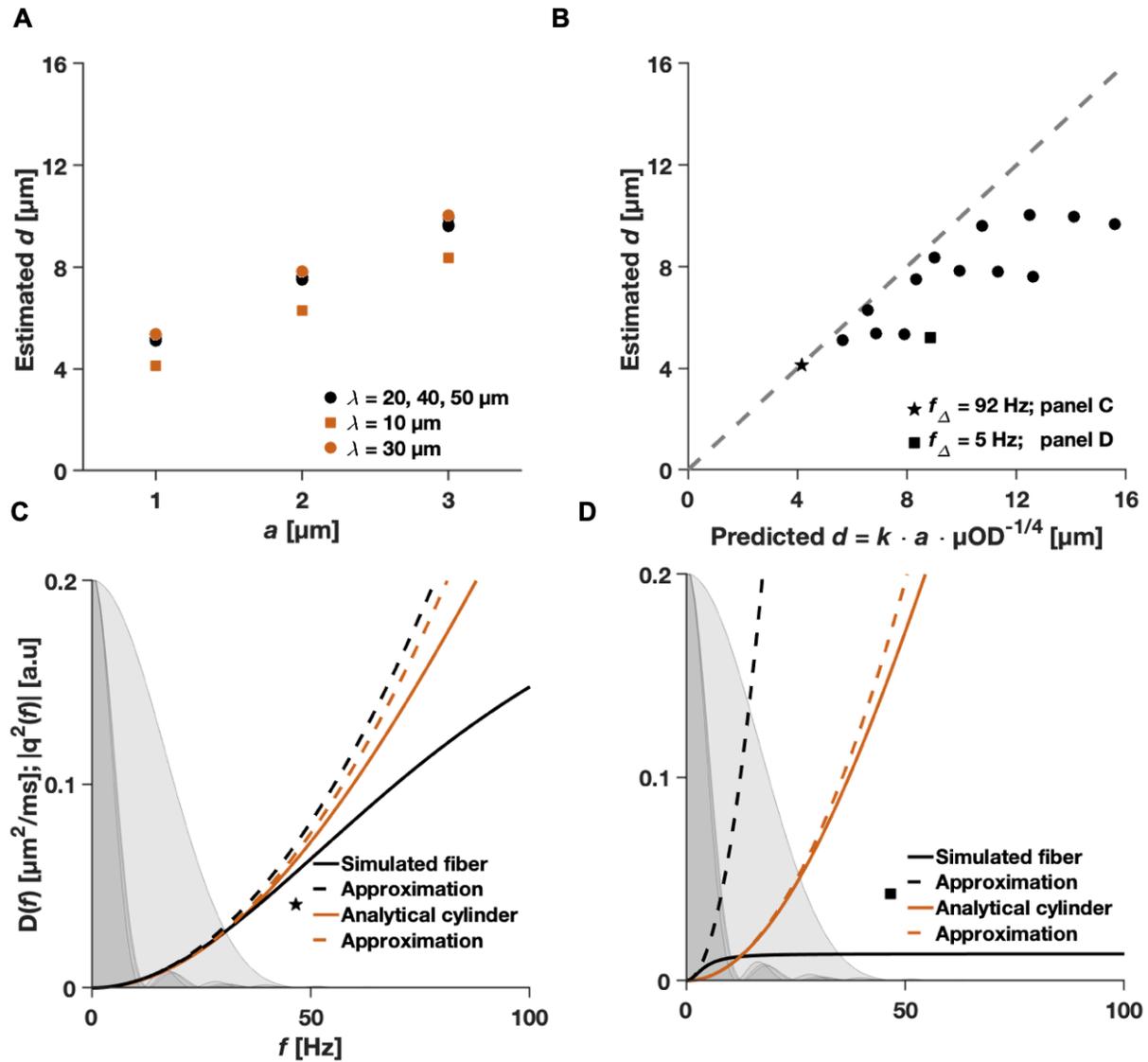

**Figure 8. The link between estimated diameters and parameters of the 1-harmonic case of the undulating thin-fiber model.** Panel A shows strong correlation between the undulation amplitude and the estimated diameter. Panel B shows predicted and estimated cylinder diameter based on a second-order Taylor expansion (Eq. 26). The predictions are more accurate for cases that had higher spectral widths (e.g. star marker, corresponding to $a = 1$ μm, $\lambda = 10$ μm and $f_\Delta = 92$ Hz) whereas those that had lower widths (e.g. square marker, $a = 1$ μm, $\lambda = 50$ μm, $f_\Delta = 5$ Hz) are underestimated. Panel C shows the case marked with star from panel B. Simulated diffusion spectrum (black solid line) of the fiber is well approximated by its second-order approximation (dashed black line, Eq. 25). Analytical diffusion spectrum corresponding to a cylinder with the estimated diameter (solid red line) is also well approximated (dashed red line, Eq. 24). Gray areas show re-scaled encoding power spectra (Table 3, Figure 2). Panel D shows the case marked with a square point from panel B where the second-order approximation of the simulated diffusion spectrum for the fiber is not a sufficient approximation.

## Tables

**Table 1. Estimated undulating parameters from histology.** Estimated values of undulation amplitude $a$, wavelength $\lambda$ and mean axon diameter $d$ of optic nerve [42 43] corpus callosum [1 39-41 46] and phrenic nerve [25 44]. The axon diameters in corpus callosum are in range 0.5-15 μm and their volume-weighted average is below 1 μm.

|  | Optic nerve | Corpus Callosum | Phrenic nerve |
|---|---|---|---|
| **Amplitude $a$ [μm]** | 5 | 1-10 | 20-100 |
| **Wavelength $\lambda$ [μm]** | 20-30 | 10-35 | 100-500 |
| **Diameter $d$ [μm]** | 1 | 0.5 | 4-5 |

**Table 2. μOD values corresponding to 1-harmonic axonal trajectories.** One μOD value (Eq. 8) can correspond to more pairs of undulation amplitudes and wavelengths.

| μOD | | Wavelength $\lambda$ [μm] | | | | |
|---|---|---|---|---|---|---|
| | | 10 | 20 | 30 | 40 | 50 |
| **Amplitude $a$ [μm]** | 1 | 0.16 | 0.05 | 0.02 | 0.01 | 0.008 |
| | 2 | 0.41 | 0.16 | 0.08 | 0.05 | 0.03 |
| | 3 | 0.59 | 0.29 | 0.16 | 0.10 | 0.07 |

**Table 3. Parameters of the pulsed diffusion gradients.** Protocol from Alexander et al. [8].

|  | $|G|$ [mTm$^{-1}$] | $\delta$ [ms] | $\Delta$ [ms] | $b$ [ms/μm$^2$] |
|---|---|---|---|---|
| **1st** | 58 | 12 | 80 | 0.50 |
| **2nd** | 46 | 15 | 77 | 0.68 |
| **3rd** | 57 | 5 | 87 | 2.45 |
| **4th** | 60 | 13 | 20 | 2.64 |

**Table 4. Comparison of mean squared errors (MSE) of the simulated and simplified spectra by the single Lorentzian approximations (Eq. 22).** Signals were generated from simulated diffusion spectra corresponding to cylinders and three examples of different cases with similar spectral width and height. They were simplified (fitted) by a single Lorentzian approximation (Figure 5, Eq. 22). The smallest error is found in the example of 1-harmonic case, the largest for a cylinder but all values are in these examples below MSE obtained for noise ($4 \cdot 10^{-4}$) at SNR = 50.

|  | Cylinders | 1-harmonic | n-harmonic | stochastic |
|---|---|---|---|---|
| **MSE of $S(f)$** | $3.1 \cdot 10^{-4}$ | $1.8 \cdot 10^{-8}$ | $1.3 \cdot 10^{-4}$ | $2.3 \cdot 10^{-4}$ |